\documentclass[showpacs,preprintnumbers,amsmath,amssymb]{revtex4}

\usepackage{epsfig}
\usepackage{graphicx}
\usepackage{dcolumn}
\usepackage{bm}

\begin{document}

\preprint{BARI-TH 482/04 }

\title{

Extending Granger causality to nonlinear systems}
\author{Nicola Ancona$^1$, Daniele Marinazzo$^{2,3}$, Sebastiano Stramaglia$^{2,3,4}$}

 \affiliation{
$^1$ Istituto di Studi sui Sistemi Intelligenti per l'Automazione, C.N.R., Bari, Italy,\\
 $^2$TIRES-Center
of
Innovative Technologies for Signal Detection and Processing,\\
Universit\`a di Bari, Italy\\
$^3$ Dipartimento Interateneo di Fisica, Bari, Italy \\
$^4$Istituto Nazionale di Fisica Nucleare, Sezione di Bari, Italy  }

\date{\today}

\begin{abstract}
We consider extension of Granger causality to nonlinear bivariate time series. In this
frame, if the prediction error of the first time series is reduced by including
measurements from the second time series, then the second time series is said to have a
causal influence on the first one. Not all the nonlinear prediction schemes are suitable
to evaluate causality, indeed not all of them allow to quantify how much the knowledge of
the other time series counts to improve prediction error. We present a novel approach
with bivariate time series  modelled by a generalization of radial basis functions and
show its application to a pair of unidirectionally coupled chaotic maps and to a
physiological example.

\end{abstract}

\maketitle
\section{Introduction\label{intro}}
Identifying causal relations among simultaneously acquired signals is an important
problem in computational time series analysis and has applications in economy [1-2], EEG
analysis \cite{tass}, human cardiorespiratory system \cite{cardioresp}, interaction
between heart rate and systolic arterial pressure \cite{akselrod}, and many others.
Several papers dealt with this problem relating it to the identification of
interdependence in nonlinear dynamical systems \cite{arnold}, or to estimates of
information rates \cite{schreiber,palus}. Some approaches modelled data by oscillators
and concentrated on the phases of the signals \cite{rosemblum}. One major approach to
analyze causality between two time series is to examine if the prediction of one series
could be improved by incorporating information of the other, as proposed by Granger
\cite{granger} in the context of linear regression models of stochastic processes. In
particular, if the prediction error of the first time series is reduced by including
measurements from the second time series in the linear regression model, then the second
time series is said to have a causal influence on the first time series. By exchanging
roles of the two time series, one can address the question of causal influence in the
opposite direction. It is worth stressing that, within this definition of causality, flow
of time plays a major role in making inference, from time series data, depending on
direction. Since Granger causality was formulated for linear models, its application to
nonlinear systems may not be appropriate. The question we address in this paper is: how
is it possible to extend Granger causality definition to nonlinear problems?

In the next section we review the original approach by Granger while describing our point
of view about its nonlinear extension; we also propose a method, exploiting radial basis
functions, which fulfills the requirements a prediction scheme should satisfy to analyze
causality. In section (\ref{exp}) we show application of the proposed method to simulated
and real examples. Some conclusions are drawn in Section (\ref{conc}).

\section{Granger causality\label{granger}}
 \subsection{Linear modelling of bivariate time series. \label{linear}}
We briefly recall the Vector AutoRegressive (VAR) model which is used to define linear
Granger causality \cite{granger}. Let $\{\bar{x}_i\}_{i=1,.,N}$ and
$\{\bar{y}_i\}_{i=1,.,N}$ be two time series of $N$ simultaneously measured quantities.
In the following we will assume that time series are stationary. For $k=1$ to $M$ (where
$M=N-m$, $m$ being the order of the model), we denote $x^k=\bar{x}_{k+m}$,
$y^k=\bar{y}_{k+m}$, ${\bf X}^k=(\bar{x}_{k+m-1}, \bar{x}_{k+m-2},...,\bar{x}_{k})$,
${\bf Y}^k=(\bar{y}_{k+m-1}, \bar{y}_{k+m-2},...,\bar{y}_{k})$ and we treat these
quantities as $M$ realizations of the stochastic variables ($x$, $y$, ${\bf X}$, ${\bf
Y}$). The following model is then considered \cite{nota1}:
\begin{eqnarray}
\begin{array}{l}
x={\bf W_{11}}\cdot {\bf X}+{\bf W_{12}}\cdot {\bf Y},\\
y={\bf W_{21}}\cdot {\bf X}+{\bf W_{22}}\cdot {\bf Y},
\end{array}
\label{lin-mod}
\end{eqnarray}
$\{{\bf W}\}$ being four $m$-dimensional real vectors to be estimated from data.
Application of least squares techniques yields the solutions:
\[ \left(\begin{array}{c} {\bf W_{11}}\\ {\bf W_{12}}\end{array}\right) = \left( \begin{array}{cc}
             \Sigma_{xx} & \Sigma_{xy} \\
              \Sigma_{yx} & \Sigma_{yy}
    \end{array}\right)^{-1} \left(\begin{array}{c} {\bf T_{11}}\\ {\bf T_{12}}\end{array}\right),\]
    and
\[ \left(\begin{array}{c} {\bf W_{21}}\\ {\bf W_{22}}\end{array}\right) = \left( \begin{array}{cc}
             \Sigma_{xx} & \Sigma_{xy} \\
              \Sigma_{yx} & \Sigma_{yy}
    \end{array}\right)^{-1} \left(\begin{array}{c} {\bf T_{21}}\\ {\bf
    T_{22}}\end{array}\right),\]
where $\Sigma$ matrices and $T$ vectors are the estimates, based on the data set at hand,
of the following average values:
\begin{eqnarray}
\begin{array}{lll}
\left[\Sigma_{xx}\right]_{\alpha\beta}=\langle X_\alpha X_\beta\rangle &={1\over M}\sum_{k=1}^M X_\alpha^k X_\beta^k & \alpha,\beta=1,...,m\\
\left[\Sigma_{xy}\right]_{\alpha\beta}=\langle X_\alpha Y_\beta\rangle&={1\over M}\sum_{k=1}^M X_\alpha^k Y_\beta^k & \alpha,\beta=1,...,m\\
\left[\Sigma_{yx}\right]_{\alpha\beta}=\langle Y_\alpha X_\beta\rangle&={1\over M}\sum_{k=1}^M Y_\alpha^k X_\beta^k & \alpha,\beta=1,...,m\\
\left[\Sigma_{yy}\right]_{\alpha\beta}=\langle Y_\alpha Y_\beta\rangle&={1\over M}\sum_{k=1}^M  Y_\alpha^k Y_\beta^k & \alpha,\beta=1,...,m\\
\left[{\bf T_{11}}\right]_{\alpha}=\langle x X_\alpha\rangle&={1\over M}\sum_{k=1}^M  x^k X_\alpha^k &  \alpha=1,...,m\\
\left[{\bf T_{12}}\right]_{\alpha}=\langle x Y_\alpha\rangle&={1\over M}\sum_{k=1}^M  x^k Y_\alpha^k & \alpha=1,...,m\\
\left[{\bf T_{21}}\right]_{\alpha}=\langle y X_\alpha\rangle&={1\over M}\sum_{k=1}^M  y^k X_\alpha^k & \alpha=1,...,m\\
\left[{\bf T_{22}}\right]_{\alpha}=\langle y Y_\alpha\rangle&={1\over M}\sum_{k=1}^M  y^k
Y_\alpha^k & \alpha=1,...,m

\end{array}
\end{eqnarray}
Let us call $\epsilon_{xy}$ and $\epsilon_{yx}$ the prediction errors of this model,
defined as the estimated variances of $x-{\bf W_{11}}\cdot {\bf X}-{\bf W_{12}}\cdot {\bf
Y}$ and $y-{\bf W_{21}}\cdot {\bf X}-{\bf W_{22}}\cdot {\bf Y}$ respectively. In
particular
\begin{eqnarray}
\begin{array}{l}
\epsilon_{xy}={1\over M}\sum_{k=1}^M (x^k-{\bf W_{11}}\cdot {\bf X}^k-{\bf W_{12}}\cdot
{\bf
Y}^k)^2;\\
\epsilon_{yx}={1\over M}\sum_{k=1}^M (y^k-{\bf W_{21}}\cdot {\bf X}^k-{\bf W_{22}}\cdot
{\bf Y}^k)^2.
\end{array}
\end{eqnarray}
We also consider AutoRegressive (AR) predictions of the two time series, i.e. the model
\begin{eqnarray}
\begin{array}{l}
x={\bf V_1}\cdot {\bf X},\\
y={\bf V_2}\cdot {\bf Y}.
\end{array}
\end{eqnarray}
In this case the least squares approach provides ${\bf V_1}=\Sigma_{xx}^{-1} {\bf
T_{11}}$ and ${\bf V_2}=\Sigma_{yy}^{-1} {\bf T_{22}}$. The estimate of the variance of
$x-{\bf V_1}\cdot {\bf X}$ is called $\epsilon_{x}$ (the prediction error when $x$ is
predicted solely on the basis of the knowledge of its past values); similarly
$\epsilon_{y}$ is the variance of $y-{\bf V_2}\cdot {\bf Y}$. If the prediction of $x$
improves by incorporating the past values of $\{y_i\}$, i.e. $\epsilon_{xy}$ is smaller
than $\epsilon_{x}$, then $y$ has a causal influence on $x$. Analogously, if
$\epsilon_{yx}$ is smaller than $\epsilon_{y}$, then $x$ has a causal influence on $y$.
Calling $c_1=\epsilon_{x}-\epsilon_{xy}$ and $c_2=\epsilon_{y}-\epsilon_{yx}$, a
directionality index can be introduced:
\begin{equation}
D={c_2 -c_1\over c_1 +c_2}.
\end{equation}
The index $D$ varies from $1$ in the case of unidirectional influence ($x\to y$) to $-1$
in the opposite case ($y\to x$), with intermediate values corresponding to bidirectional
influence. According to this definition of causality, the following property holds for
$N$ sufficiently large: {\it if ${\bf Y}$ is uncorrelated with ${\bf X}$ and $x$, then
$\epsilon_{x}=\epsilon_{xy}$}. Indeed in this case $\Sigma_{xy}=\Sigma_{yx}=0$ and ${\bf
T_{12}}=0$, therefore ${\bf W_{12}}=0$. This means that VAR and AR modelling of the
$\{x_i\}$ time series coincide. Analogously {\it if ${\bf X}$ is uncorrelated with ${\bf
Y}$ and $y$, then $\epsilon_{y}=\epsilon_{yx}$}. It is clear that these properties are
fundamental and make the linear prediction approach suitable to evaluate causality.  On
the other hand, for nonlinear systems higher order correlations may be relevant.
Therefore, we propose that any prediction scheme providing a nonlinear extension of
Granger causality should satisfy the following property: (P1) {\it if ${\bf Y}$ is
statistically independent of ${\bf X}$ and $x$, then $\epsilon_{x}=\epsilon_{xy}$}; {\it
if ${\bf X}$ is statistically independent of ${\bf Y}$ and $y$, then
$\epsilon_{y}=\epsilon_{yx}$}. In a recent paper \cite{chen}, use of a locally linear
prediction scheme \cite{farmer} has been proposed to evaluate nonlinear causality. In
this scheme, the joint dynamics of the two time series is reconstructed by delay vectors
embedded in an Euclidean space; in the delay embedding space a locally linear model is
fitted to data. The approach described in \cite{chen} satisfies property P1 only if the
number of points in the neighborhood of each reference point, where linear fit is done,
is sufficiently high to establish good statistics; however linearization is valid only
for small neighborhoods. It follows that this approach to nonlinear causality requires
very long time series to satisfy P1. In order to construct methods  working effectively
with moderately long time series, in the next subsection we will characterize the problem
of extending Granger causality as the one of finding classes of nonlinear models
satisfying property P1.

\subsection{Nonlinear models. \label{nonlinear}}
What is the most general class of nonlinear models which satisfy P1? The complete answer
to this question is matter for further study. Here we only give a partial answer, i.e.
the following family of models:
\begin{eqnarray}
\begin{array}{l}
x={\bf w_{11}}\cdot {\bf \Phi}\left({\bf X}\right)+{\bf w_{12}}\cdot {\bf \Psi}\left({\bf Y}\right),\\
y={\bf w_{21}}\cdot {\bf \Phi}\left({\bf X}\right)+{\bf w_{22}}\cdot {\bf \Psi}\left({\bf
Y}\right), \label{mod-non}
\end{array}
\end{eqnarray}
where $\{\bf w\}$ are four $n$-dimensional real vectors, ${\bf
\Phi}=\left(\varphi_1,...,\varphi_n\right)$ are $n$ given nonlinear real functions of $m$
variables, and ${\bf \Psi}=\left(\psi_1,...,\psi_n\right)$ are $n$ other  real functions
of $m$ variables. Given ${\bf \Phi}$ and ${\bf \Psi}$, model (\ref{mod-non}) is a linear
function in the space of features $\varphi$ and $\psi$; it depends on $4n$ variables, the
vectors $\{\bf w\}$, which must be fixed to minimize the prediction errors
\begin{eqnarray}
\begin{array}{l}
\epsilon_{xy}={1\over M}\sum_{k=1}^M (x^k-{\bf w_{11}}\cdot {\bf \Phi}\left({\bf X}^k\right)-{\bf w_{12}}\cdot {\bf \Psi}\left({\bf Y}^k\right))^2;\\
\epsilon_{yx}={1\over M}\sum_{k=1}^M (y^k-{\bf w_{21}}\cdot {\bf \Phi}\left({\bf
X}^k\right)-{\bf w_{22}}\cdot {\bf \Psi}\left({\bf Y}^k\right))^2.
\end{array}
\end{eqnarray}

We also consider the model:
\begin{eqnarray}
\begin{array}{l}
x={\bf v_{1}}\cdot {\bf \Phi}\left({\bf X}\right),\\
y={\bf v_{2}}\cdot {\bf \Psi}\left({\bf Y}\right),
\end{array}
\label{mmod}
\end{eqnarray}
and the corresponding prediction errors $\epsilon_{x}$ and $\epsilon_{y}$.

Now we prove that model (\ref{mod-non}) satisfies P1. Let us suppose that ${\bf Y}$ is
statistically independent of ${\bf X}$ and $x$. Then, for each $\mu=1,..,n$ and for each
$\lambda=1,..,n$: $\psi_\mu \left({\bf Y}\right)$ is uncorrelated with $x$ and with
$\varphi_\lambda \left({\bf X}\right)$. It follows that
\begin{equation}
\mbox{variance}\left[x-{\bf w_{11}}\cdot {\bf \Phi}\left({\bf X}\right)-{\bf w_{12}}\cdot
{\bf \Psi}\left({\bf Y}\right)\right]=\mbox{variance}\left[x-{\bf w_{11}}\cdot {\bf
\Phi}\left({\bf X}\right)\right]+\mbox{variance}\left[{\bf w_{12}}\cdot {\bf
\Psi}\left({\bf Y}\right)\right].
\end{equation}
As a consequence, for large $N$, at the minimum of $\epsilon_{xy}$ one has ${\bf
w_{12}}=0$. The same argument may be used exchanging x and y. This proves that P1 holds.

The solution of least squares fitting of model (\ref{mod-non}) to data may be written in
the following form:
\[ \left(\begin{array}{c} {\bf w_{11}}\\ {\bf w_{12}}\end{array}\right) = \left( {\bf S_{1}}\;\;\; {\bf S_{2}}\right)^\dag {\bf t_1},\]

\[ \left(\begin{array}{c} {\bf w_{21}}\\ {\bf w_{22}}\end{array}\right) = \left( {\bf S_{2}}\;\;\; {\bf S_{1}}\right)^\dag {\bf t_2},\]

where $^\dag$ denotes the pseudo-inverse matrix \cite{rao}; ${\bf S}$ matrices and ${\bf
t}$ vectors are given by:
\begin{eqnarray}
\begin{array}{ll}
\left[{\bf S_{1}}\right]_{k \rho}=\varphi_\rho\left({\bf X}^k\right) & k=1,...,M, \rho=1,...,n\\
\left[{\bf S_{2}}\right]_{k \rho}=\psi_\rho({\bf Y}^k) & k=1,...,M, \rho=1,...,n\\
\left[{\bf t_{1}}\right]_{k}=x^k & k=1,...,M\\
\left[{\bf t_{2}}\right]_{k}=y^k & k=1,...,M
\end{array}
\end{eqnarray}
Solution of model (\ref{mmod}) is given by ${\bf v_{1}}={\bf S_1}^\dag{\bf t_1}$ and
${\bf v_{2}}={\bf S_2}^\dag {\bf t_2}$.
\subsection{Radial basis functions. \label{rbf}}
Radial basis functions (RBF) methods were initially proposed to perform exact
interpolation of a set of data points in a multidimensional space (see, e.g.,
\cite{bishop}); subsequently an alternative motivation for RBF methods was found within
regularization theory \cite{poggio}. RBF models have been used to model financial time
series \cite{hutch}.

In this subsection we propose a strategy to choose the functions ${\bf \Phi}$ and ${\bf
\Psi}$, in model (\ref{mod-non}), in the frame of RBF methods. Fixed $n\ll M$, $n$
centers $\{{\bf \tilde{X}}^\rho\}_{\rho=1}^n$, in the space of ${\bf X}$ vectors, are
determined by a clustering procedure applied to data $\{{\bf X}^k\}_{k=1}^M$. Analogously
$n$ centers $\{{\bf \tilde{Y}}^\rho\}_{\rho=1}^n$, in the space of ${\bf Y}$ vectors, are
determined by a clustering procedure applied to data $\{{\bf Y}^k\}_{k=1}^M$. We then
make the following choice:
\begin{eqnarray}
\begin{array}{ll}
\varphi_\rho
\left({\bf X}\right)=\exp\left({-\|{\bf X}-{\bf \tilde{X}}^\rho\|^2/ 2\sigma^2}\right)&\rho=1,...,n,\\
\psi_\rho \left({\bf Y}\right)=\exp\left({-\|{\bf Y}-{\bf \tilde{Y}}^\rho\|^2 /
2\sigma^2}\right)&\rho=1,...,n,
\end{array}
\label{eq-rbf}
\end{eqnarray}
$\sigma$ being a fixed parameter, whose order of magnitude is the average spacing between
the centers. Centers $\{{\bf \tilde{X}}^\rho\}$ are the prototypes  of ${\bf X}$
variables, hence $\varphi$ functions measure the similarity to these typical patterns.
Analogously, $\psi$ functions measure the similarity to typical patterns of ${\bf Y}$.
Many clustering algorithm may be applied to find prototypes, for example in our
experiments we use fuzzy c-means \cite{fcm}.

Some remarks are in order. First, we observe that the models  described above may
trivially be adapted to handle the case of reconstruction embedding of the two time
series in a delay coordinate space, as described in \cite{chen}. Second, we stress that
in (\ref{mod-non}) $x$ and $y$ are modelled as the sum of two contributions, one
depending solely on ${\bf X}$ and the other dependent on ${\bf Y}$. Obviously better
prediction models for $x$ and $y$ exists, but they would not be useful to evaluate
causality unless they would satisfy P1. This requirement poses a limit to the level of
detail at which the two time series may be described, if one is looking at causality
relationships. The justification of the model we propose here, based on regularization
theory, is sketched in the Appendix.
\subsection{Empirical risk and generalization error. \label{error}}
In the previous subsections the prediction error has been identified as the empirical
risk, although there is a difference between these two quantities as Statistical Learning
Theory (SLT) \cite{vapnik-book-1998} shows. The deep connection between empirical risk
and generalization error deserves a comment here. First of all we want to point out that
the ultimate goal of a predictor and in general of any supervised machine $x=f({\bf X})$
\cite{nota2} is {\it to generalize}, that is to correctly predict the output values $x$
corresponding to never seen before input patterns ${\bf X}$ (for definiteness we consider
the case of predicting $x$ on the basis of the knowledge of ${\bf X}$). A measure of the
generalization error of such a machine $f$ is the {\it risk} $R[f]$ defined as the
expected value of the loss function $V\left( x, f({\bf X})\right)$:
\begin{equation}
R[f] = \int dx\; d{\bf X}\;\;V\left( x, f({\bf X})\right) P(x,{\bf X}) , \label{er1}
\end{equation}
\noindent where $P(x,{\bf X})$ is the probability density function underlying the data. A
typical example of loss function is $V\left( x, f({\bf X})\right) = \left( x - f({\bf
X})\right)^2$ and in this case the function minimizing $R[f]$ is called the {\it
regression function}. In general $P$ is unknown and so we can not minimize the risk. The
only data we have are $M$ observations (examples) $S=\{(x^k, {\bf X}^k)\}_{k=1}^M$ of the
random variables $x$ and ${\bf X}$ drawn according to $P(x,{\bf X})$. Statistical
learning theory \cite{vapnik-book-1998} as well as regularization theory \cite{poggio}
provide upper bounds of the generalization error of a learning machine $f$. Inequalities
of the following type may be proven:
\begin{equation}
R[f] \leq  \epsilon_{x} + {\cal C}, \label{ineq}
\end{equation}
where
\begin{equation}
\epsilon_x ={1\over M}\sum_{k=1}^M \left(x^k - f({\bf X}^k)\right)^2 \end{equation} is
the {\it empirical risk}, that measures the error on the training data. ${\cal C}$ is a
measure of the {\it complexity} of machine $f$ and it is related to the so-called
Vapnik-Chervonenkis (VC) dimension. Predictors with low complexity guarantee low
generalization error because they avoid overfitting to occur. When the complexity of the
functional space where our predictor lives is {\it small}, then the empirical risk is a
good approximation of the generalization error. The models we deal with in this work
verify such constraint. In fact, linear predictors have a finite VC-dimension, equal to
the size of the space where the input patterns live, and predictors expressed as linear
combinations of radial basis functions are smooth. In conclusion empirical risk is a good
measure of the generalization error for the predictors we are considering here and so it
can be used to construct measures of causality between time series \cite{loo}.
\section{Experiments. \label{exp}}
In order to demonstrate the use of the proposed approach, in this section we study two
examples, a pair of unidirectionally coupled chaotic maps and a bivariate physiological
time series.
\subsection{Chaotic maps. \label{chaotic}}
Let us consider the following pair of noisy logistic maps:
\begin{eqnarray}
\begin{array}{l}
x_{n+1}=a\;x_n\;\left(1-x_n\right)+s \eta_{n+1},\\
y_{n+1}=e\;a\;y_n\;\left(1-y_n\right)+(1-e)\;a\; x_n\;\left(1-x_n\right)+s \xi_{n+1};
\end{array}
\label{map}
\end{eqnarray}
$\{\eta\}$ and $\{\xi\}$ are unit variance Gaussianly distributed noise terms; parameter
$s$ determines their relevance. We fix $a=3.8$, and $e\in [0,1]$ represents the coupling
$x \to y$. In the noise-free case ($s=0$), a transition to synchronization ($x_n =y_n$)
occurs at $e=0.37$. We evaluate the Lyapunov exponents by the method described in
\cite{bremen}: the first exponent is $0.43$, the second exponent depends on $e$ and is
depicted in Fig. 1 for $e<0.37$ (it becomes negative for $e>0.37$). For several values of
$e$, we have considered runs of $10^5$ iterations, after $10^5$ transient, and evaluated
the prediction errors by (\ref{mod-non}) and (\ref{mmod}), with $m=1$, $n=100$ and
$\sigma=0.05$. In fig. 2a we depict, in the noise free case, the curves representing
$c_1$ and $c_2$ versus coupling $e$.  In figures 2b, 2c and 2d we depict the
directionality index $D$ versus $e$, in the noise free case and for $s=0.01$ and $s=0.07$
respectively. In the noise free case we find $D=1$, i.e. our method revealed
unidirectional influence. As the noise increases, also the minimum value of $e$, which
renders unidirectional coupling detectable, increases.
\subsection{Physiological data. \label{physiol}}
As a real example, we consider time series of heart rate and breath rate of a sleeping
human suffering from sleep apnea (ten minutes from data set B of the Santa Fe Institute
time series contest held in 1991, available in the Physionet data bank \cite{physionet}).
There is a growing evidence that suggests a causal link between sleep apnea and
cardiovascular disease \cite{roux}, although the exact mechanisms that underlie this
relationship remain unresolved \cite{somno}. Figure 3 clearly shows that bursts of the
patient breath and cyclical fluctuations of heart rate are interdependent. We fix $m=1$
and $n=50$; varying $\sigma$ we find that both $\epsilon_x$ ($x$ representing heart rate)
and $\epsilon_y$ ($y$ representing breath) have a minimum at $\sigma$ close to $0.5$. In
fig. 4 we depict the directionality index $D$ vs $\sigma$, around $\sigma =0.5$. Since we
find $D$ positive, we may conclude that the causal influence of heart rate on breath is
stronger than the reverse \cite{nota3}. This data have been already analyzed in
\cite{schreiber}, measuring the rate of information flow (transfer entropy), and a
stronger flow of information from the heart rate to the breath rate was found. In this
example, the rate of information flow entropy and Granger nonlinear causality give
consistent results: both these quantities, in the end, measure the departure from the
generalized Markov property \cite{schreiber}
\begin{eqnarray}
\begin{array}{ll}
P(x\;|\;{\bf X})\;=\;P(x\;|\;{\bf X},{\bf Y}),\\
P(y\;|\;{\bf Y})\;=\;P(y\;|\;{\bf X},{\bf Y}).
\end{array}
\label{markov}
\end{eqnarray}
\section{Conclusions. \label{conc}}
The components of complex systems in nature rarely display a linear interdependence of
their parts: identification of their causal relationships provides important insights on
the underlying mechanisms. Among the variety of methods which have been proposed to
handle this important task, a major approach was proposed by Granger \cite{granger}. It
is based on the improvement of predictability of one time series due to the knowledge of
the second time series: it is appealing for its general applicability, but is restricted
to linear models. While extending Granger approach to the nonlinear case, on one hand one
would like to have the most accurate modelling of the bivariate time series, on the other
hand the goal is to quantify how much the knowledge of the other time series counts to
reach this accuracy. Our analysis is rooted on the fact that any nonlinear modelling of
data, suitable to study causality, should satisfy the property P1, described in Section
(\ref{granger}). It is clear that this property sets a limit on the accuracy of the
model; we have proposed a class of nonlinear models which satisfy P1 and constructed an
RBF like approach to nonlinear Granger causality. Its performances, in a simulated case
and a real physiological application, have been presented. We conclude remarking that use
of this definition of nonlinear causality may lead to discover genuine causal structures
via data analysis, but to validate the results the analysis has to  be accompanied by
substantive theory.

\vskip 0.4 cm\par\noindent{\bf Acknoledgements.} The authors thank Giuseppe Nardulli and
Mario Pellicoro for useful discussions about causality.
\section{Appendix. \label{app1}}
We show how the choice of functions (\ref{eq-rbf}) arise in the frame of regularization
theory. Let $z$ be a function of ${\bf X}$ and ${\bf Y}$. We assume that $z$ is the sum
of a term depending solely on ${\bf X}$ and one depending on ${\bf Y}$: $z({\bf X},{\bf
Y})=f({\bf X})+g({\bf Y})$. We also assume the knowledge of the values of $f$ and $g$ at
points $\{{\bf \tilde{X}}^\rho, {\bf \tilde{Y}}^\rho\}_{\rho=1,..,n}$:
\begin{eqnarray}
\begin{array}{ll}
f({\bf \tilde{X}}^\rho)=f^\rho &\rho=1,...,n,\\
g({\bf \tilde{Y}}^\rho)=g^\rho &\rho=1,...,n.\\
\end{array}
\label{app1}
\end{eqnarray}
Let us denote $\hat{K}( \vec{\omega})$ the Fourier transform of $K(\vec{r})=\exp(-
r^2/2\sigma^2)$. The following functional is a measure of the smoothness of z({\bf
X},{\bf Y}):
\begin{equation}
{\cal S}[z]=\int
d\vec{\omega}\;\;{|\hat{f}(\vec{\omega})|^2+|\hat{g}(\vec{\omega})|^2\over \hat{K}
(\vec{\omega})}. \label{app2} \end{equation} Indeed it penalizes functions with relevant
contributions from high frequency modes. Variational calculus shows that the function
that minimize ${\cal S}$ under the constraints (\ref{app1}) is given by:
\begin{equation}
z=\sum_{\rho=1}^n \mu_\rho\; K \left( {\bf X}-{\bf \tilde{X}}^\rho\right)+\sum_{\rho=1}^n
\lambda_\rho\; K\left( {\bf Y}-{\bf \tilde{Y}}^\rho \right); \label{app3}
\end{equation}
where $\{\mu\}$ and $\{\lambda\}$ are tunable Lagrange multipliers to solve (\ref{app1}).
Hence model (\ref{mod-non})-(\ref{eq-rbf}) corresponds to the class of the smoothest
functions, sum of a term depending on ${\bf X}$ and a term depending on ${\bf Y}$, with
assigned values on a set of $n$ points.

\begin{figure}[ht!]
\begin{center}
\epsfig{file=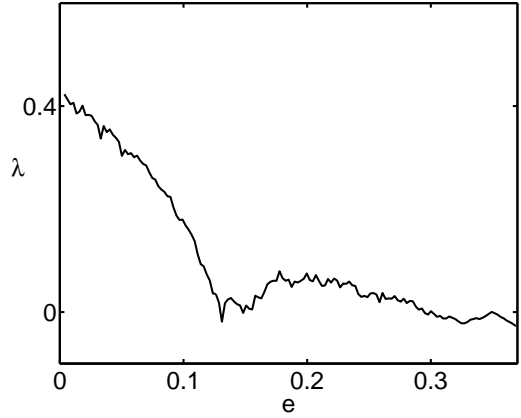,height=5.5cm}
\end{center}
\caption{{\small  The second Lyapunov exponent of the coupled maps (\ref{map}) is plotted
versus coupling $e$. \label{fig1}}}
\end{figure}

\begin{figure}[ht!]
\begin{center}
\epsfig{file=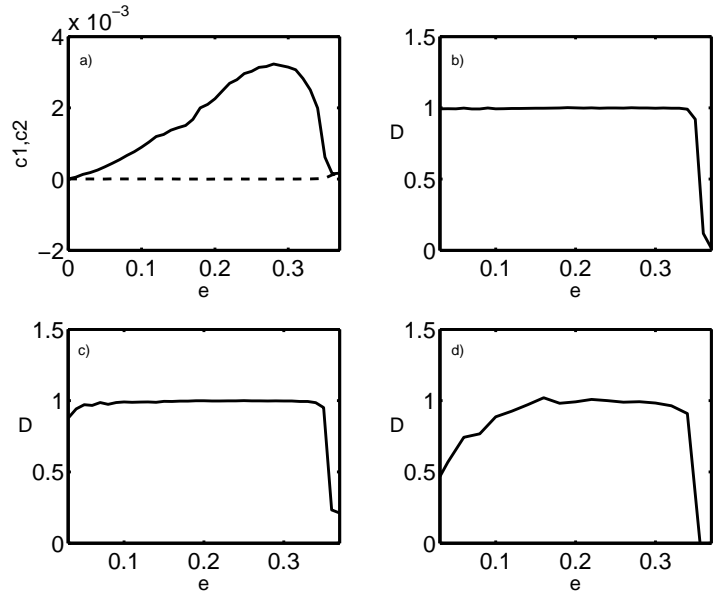,height=8.cm}
\end{center}
\caption{{\small  (a) For the noise free case of coupled maps (\ref{map}),
$c_1=\epsilon_{x}-\epsilon_{xy}$ (dashed line) and $c_2=\epsilon_{y}-\epsilon_{yx}$
(solid line) are plotted versus coupling $e$. (b) The directionality index $D$ (see the
text) is plotted versus $e$ in the noise free case. (c) The directionality index $D$ is
plotted versus $e$, $s=0.01$. (d) $D$ is plotted versus $e$, $s=0.07$. \label{fig2}}}
\end{figure}

\begin{figure}[ht!]
\begin{center}
\epsfig{file=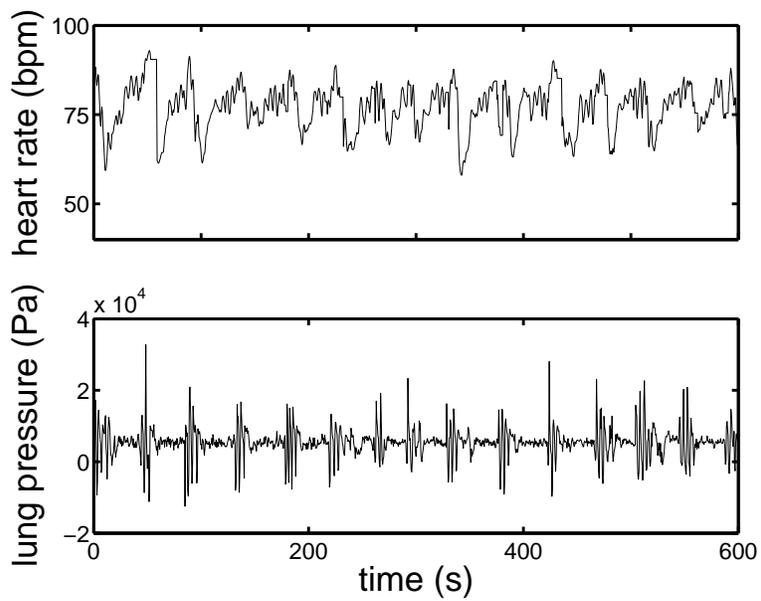,height=8.cm}
\end{center}
\caption{{\small Time series of the heart RR (upper) and breath signal (lower) of a
patient suffering sleep apnea. Data sampled at 2 Hz. \label{fig3}}}
\end{figure}

\begin{figure}[ht!]
\begin{center}
\epsfig{file=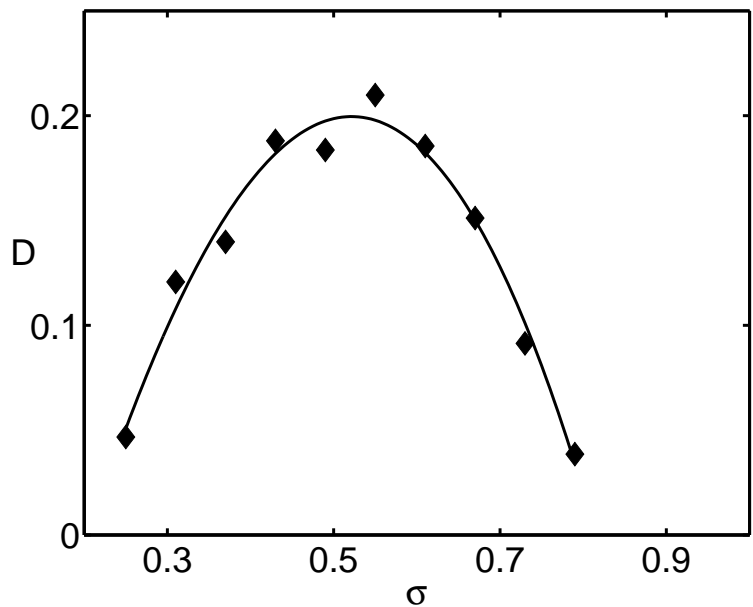,height=8.cm}
\end{center}
\caption{{\small The directionality index $D$ is plotted versus $\sigma$ for the
physiological application, around $\sigma =0.5$. Solid line is the $3th$-polynomial best
fit of points, here shown only for illustrative purposes.  \label{fig4}}}
\end{figure}


\begin{thebibliography}{99}
\bibitem{nota1} Usually both times series are normalized in the preprocessing stage,
i.e. they are linearly transformed to have zero mean and unit variance.
\bibitem{loo}In the general case, Leave-one-out (Loo) error $E_{loo}[f] $  provides
a better estimate of the generalization error of $f$ (Luntz and Brailovsky theorem) than
the empirical risk, given a finite number of training data. Loo-error is defined as the
error variance when the prediction for the k-th pattern is made using the model trained
on the M-1 other patterns; it needs $M$ predictors to be trained, where $M$ is the
cardinality of the data set. Hence,  Loo-error estimation is unfeasible to compute for
large training sets. For linear predictors, like the ones we consider in this paper, the
empirical risk is already a good estimate, due to the low complexity of these machines.
\bibitem{nota3} These results may also be due to coupling of the two signals to
a common external driver.
\end{thebibliography}
\end{document}